\documentclass[conference]{IEEEtran}

\usepackage[english]{babel}
\usepackage{float}
\usepackage{amsmath}
\usepackage{graphicx}
\usepackage[colorlinks=true, allcolors=blue]{hyperref}

\title{Reconfigurable, large-format D-ToF/photon-counting SPAD image sensors with embedded FPGA for scene adaptability}
\author{
    \IEEEauthorblockN{Tommaso Milanese\IEEEauthorrefmark{1}, Baris Can Efe\IEEEauthorrefmark{1}, Claudio Bruschini\IEEEauthorrefmark{1}, Nobukazu Teranishi\IEEEauthorrefmark{2}, Edoardo Charbon\IEEEauthorrefmark{1}}
    \IEEEauthorblockA{\IEEEauthorrefmark{1}AQUA, STI School of engineering, École polytechnique fédérale de Lausanne, Neuchâtel CH-2000, Switzerland\\\IEEEauthorrefmark{2}Research Institute of Electronics, Shizuoka University, Hamamatsu, Japan
    \\\{tommaso.milanese, baris.efe, claudio.bruschini, edoardo.charbon\}@epfl.ch, {teranishi}@idl.rie.shizuoka.ac.jp}
}

\begin{document}
\setlength{\abovedisplayskip}{10pt}
\setlength{\belowdisplayskip}{10pt}
\setlength{\abovedisplayshortskip}{0pt}
\setlength{\belowdisplayshortskip}{0pt}
\maketitle

\begin{abstract}
CMOS-compatible single-photon avalanche diodes (SPADs) have emerged in many systems as the solution of choice for cameras with photon-number resolution and photon counting capabilities. Being natively digital optical interfaces, SPADs are naturally drawn to {\em in situ} logic processing and event-driven computation; they are usually coupled to discrete FPGAs to enable reconfigurability. In this work, we propose to bring the FPGA on-chip, in direct contact with the SPADs at pixel or cluster level. To demonstrate the suitability of this approach, we created an architecture for processing timestamps and photon counts using programmable weighted sums based on an efficient use of look-up tables. The outputs are processed hierarchically, similarly to what is done in FPGAs, reducing power consumption and simplifying I/Os. Finally, we show how artificial neural networks can be designed and reprogrammed by using look-up tables in an efficient way.
\end{abstract}

\section{Introduction}
SPADs are optical devices capable of detecting and counting single photons, ideally suited for digital processing. Unlike conventional photodiodes integrated in imaging devices, SPADs offer a direct digital output, whereas no A/D conversion is needed. Their potential has been exploited by major industries using standard CMOS processes \cite{pellegrini2017industrialised}, transitioning from custom technologies for niche applications to mass-production \cite{stm:vl5}.
In this work, we introduce a novel architecture for the processing and readout of SPAD sensors, integrating a field-programmable gate array (FPGA) fabric on chip. This approach introduces several key innovations in both SPAD front-end and overall array architecture, leading to the following key contributions:
\begin{itemize}
\item A reconfigurable SPAD macropixel optimized for photon counting and Time-of-Flight (ToF) applications;
\item A reconfigurable Region-of-Interest (ROI) across the SPAD array, enabling low-power operation modes;
\item Continuous-time embedded neuromorphic computing utilizing SPAD digital pulses.
\end{itemize}
Previous works have explored different SPAD macropixel configurations for digital silicon photomultipliers (dSiPMs), employing fixed SPAD combinations to enhance performance in particular applications. In \cite{6154110} the authors propose a spatio-temporal compression circuit based on a monostable circuit and an OR tree. This method provides a high-dynamic range while keeping an event-driven architecture for timestamping, despite the fact that the dynamic range is limited by the monostable circuit and the number of SPADs. 
Another method is introduced in \cite{dutton201511} and expanded in \cite{hutchings2019reconfigurable}, where the SPADs are connected to a XOR tree with a toggle flip-flop (TFF) redirecting its output to a time-to-digital converter (TDC), which timestamps photons in both rising and falling edges of the XOR tree. This readout technique, however, is limited in terms of resolution and TDC range. In \cite{huang2023dual} a dual readout mode is proposed for a dSiPM, incorporating both single-hit and multi-hit readout of the dSiPM through custom CMOS logic, selectable during chip operation by the host. 
\begin{figure}[h!]
    \centering
    \includegraphics[trim= 1 1 1 1,clip,width=0.9\linewidth]{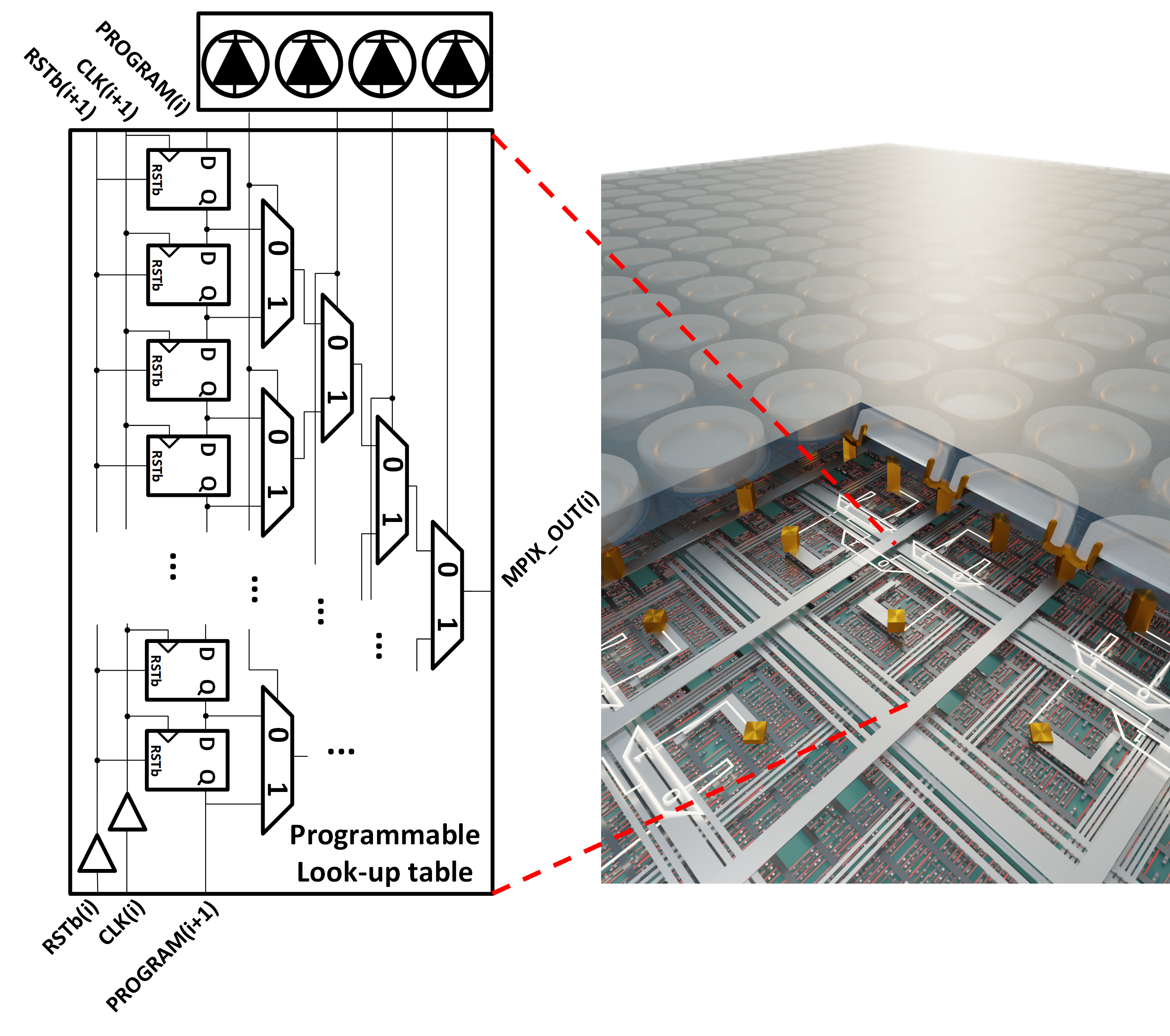}
    \caption{Concept proposed in this paper. A 2$\times$2 SPAD array is coupled to a linear array of memories, which defines how the output of each SPAD is combined with one another, so as to achieve a reconfigurable tree that can be combined with other trees in the imager (Drawing: Harald Homulle).}
    \vspace{-0.1in}
    \label{fig_system_FPGA_archi}
\end{figure}
In \cite{ximenes} SPADs are organized in a cluster and connected through a tree of decision-making components that redirect a traveling pulse to a TDC, preserving the origin of the pulse in a winner-take-all format for each cluster. This technique enables sharing a TDC per cluster while maintaining high timing resolution, but, like the XOR technique, it is fixed. In \cite{milanese2023linospad2} a SPAD linear array is connected to an FPGA that can implement a large number of functions, including an array of TDCs, but it requires, like any FPGAs, a large overhead and high power consumption.

In this work, we introduce a reconfigurable SPAD readout based on look-up tables (LUTs), optimizing photon-counting performance through XOR combinations, and ToF measurements via OR pulse combinations and multi-level coincidences. Fig.~\ref{fig_system_FPGA_archi} shows the principle, whereas in this case SPADs are in the top tier of a 3D-stacked chip and the processing electronics is in the bottom tier. We demonstrate this concept with a monolithic test chip implemented in a 110 nm CIS process, where 16 SPADs are hierarchically combined using a two-level LUT architecture. Furthermore, we extend programmability at the SPAD data readout level, allowing selection between counting LUT pulses or timestamping individual pulses. This flexibility enables Region-of-Interest (ROI) selection within the array for low-power operation. In addition, we show a neuromorphic computing application, again relying on the combination of SPADs by means of LUTs. 
\section{Reconfigurable SPAD architecture}
We explored the proposed architecture implemented in a 110 nm CIS process to perform different functions. 
\subsection{Pixel circuit}
\begin{figure}[h!]
    \centering
    \includegraphics[width=\linewidth]{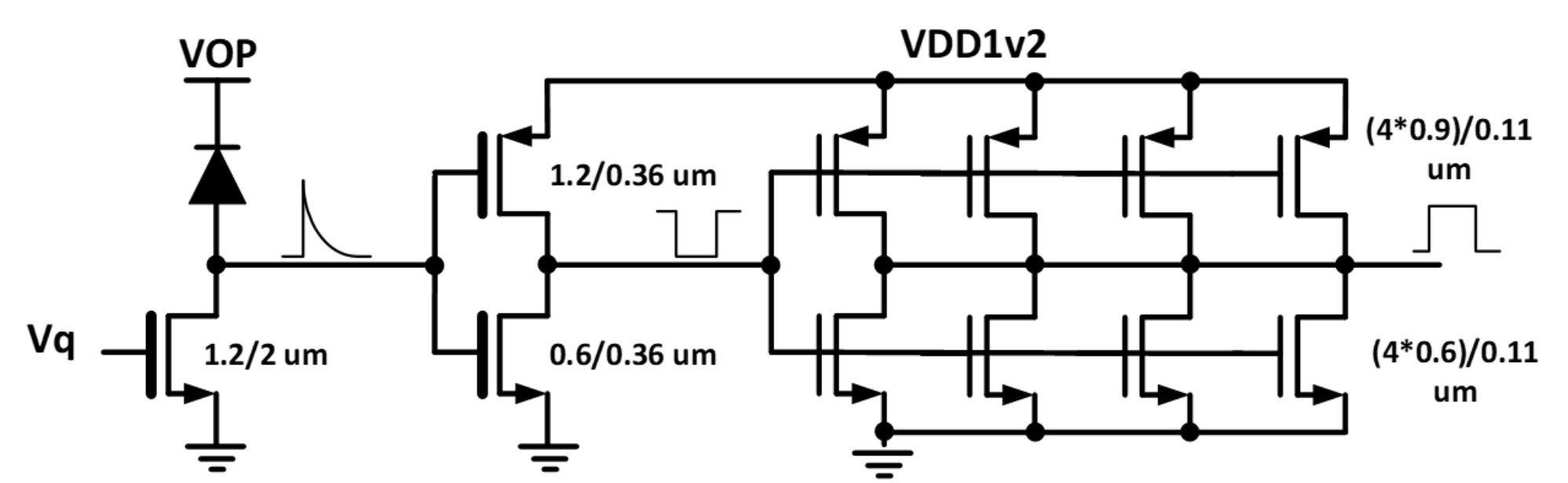}
    \caption{Pixel circuit schematic. The SPAD is quenched with a thick oxide transistor and the pulse is discretized/level shifted by the output buffer.}
    \label{fig_pix_circuit}
\end{figure}

The pixel circuit comprises a passive quench and recharge front-end with a buffer/level shifter for 1.2 V operation. The circuit schematic is shown in Fig.~\ref{fig_pix_circuit}.
A $\mathrm{W}/\mathrm{L}<1$ ratio for the quenching transistor has been chosen to enable a wide range of equivalent quenching resistances according to the $\mathrm{V}_{q}$ quenching voltage. This translates to a wide range of SPAD dead times, in order to test the LUT combinations by varying the SPAD pulse width. Larger dimensions than minimum sizes have also been employed to lower fabrication variabilities.
The CMOS 1.2 V digital pulse is sent for further processing to the LUTs. 

\subsection{Look-up table}
\begin{figure}[h!]
    \centering
    \includegraphics[trim= 1 10 1 10,clip,width=1\linewidth]{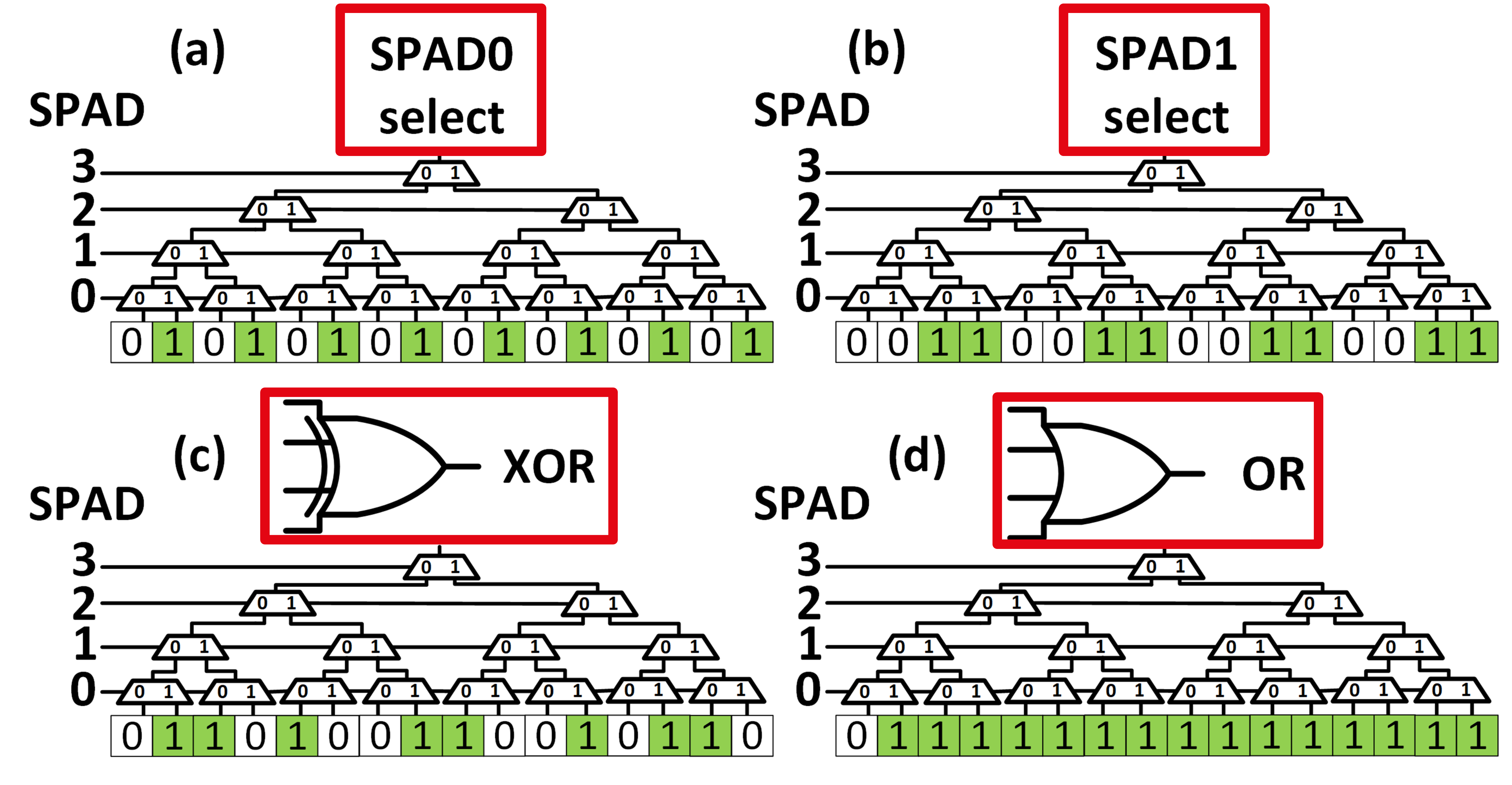}
    \caption{Examples of 4 SPAD readout combinations enabled by the LUT.}
    \label{fig_SPADcombexample}
\end{figure}

\begin{figure}[h!]
    \centering
    \includegraphics[width=1\linewidth]{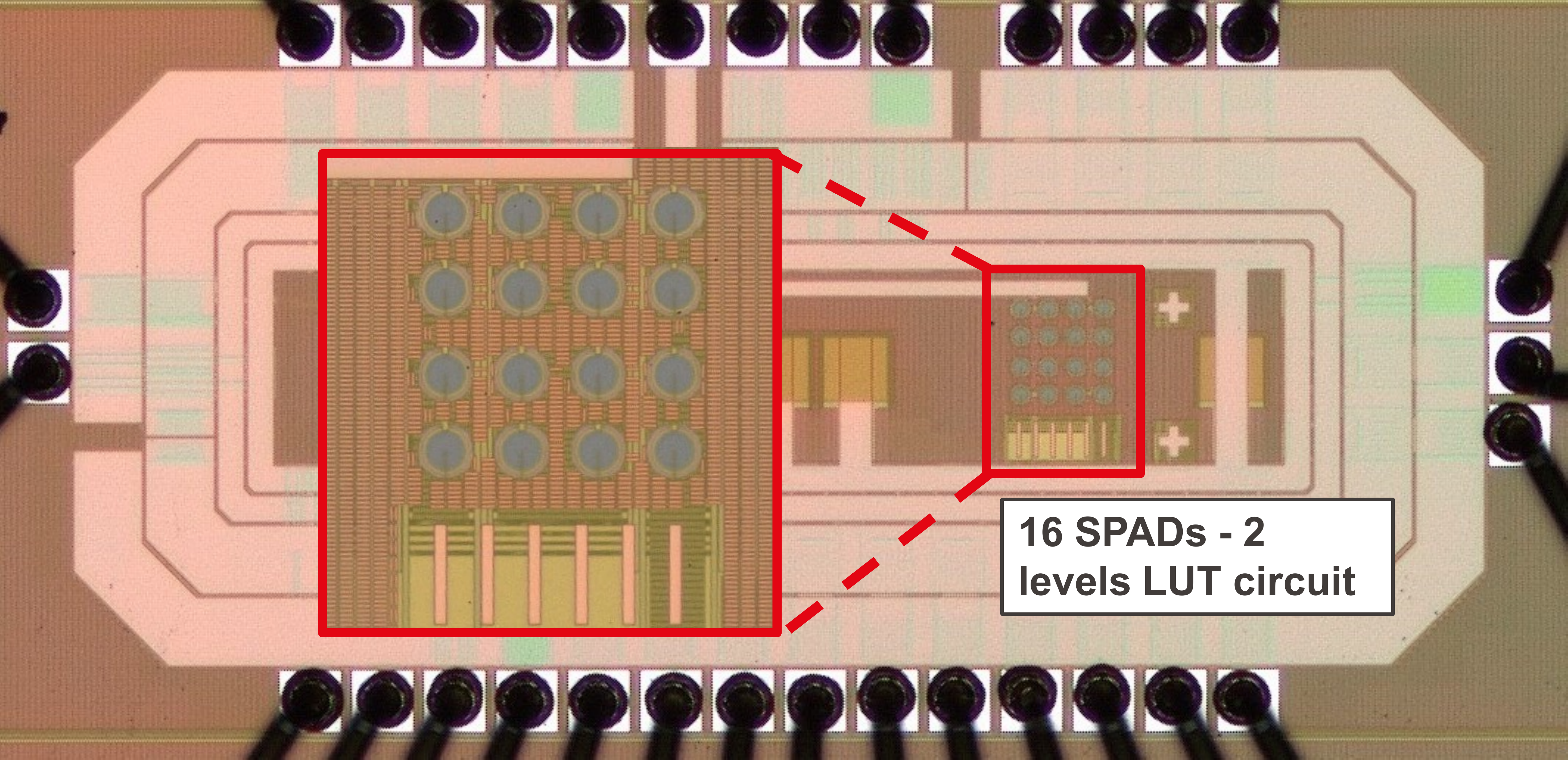}
    \caption{Test chip micrograph fabricated in 110 nm CIS technology. The 16 SPADs macropixel is highlighted in the figure.}
    \vspace{-0.1in}
    \label{fig_ugraph}
\end{figure}

The LUT is implemented using standard cells operating at 1.2 V and programmed through a serial interface using a long shift register. 
The chip comprises 5 LUTs with 4 entries per LUT (4 SPAD pixels), which is equivalent to 80 flip flops, 16 per LUT, as shown in Fig.~\ref{fig_system_FPGA_archi}. Examples of the possible SPAD readout configurations are shown in Fig.~\ref{fig_SPADcombexample}.
The SPADs comprise a isolated P+/Deep Nwell junction; their layout is shown in the micrograph of Fig.~\ref{fig_ugraph}.
With a clock of 100 MHz, the chip programming time is 800 ns,
fast enough to reconfigure the imager in real time.

\subsection{Neuromorphic computing}
\begin{figure}[h!]
    \centering
    \includegraphics[width=0.8\linewidth]{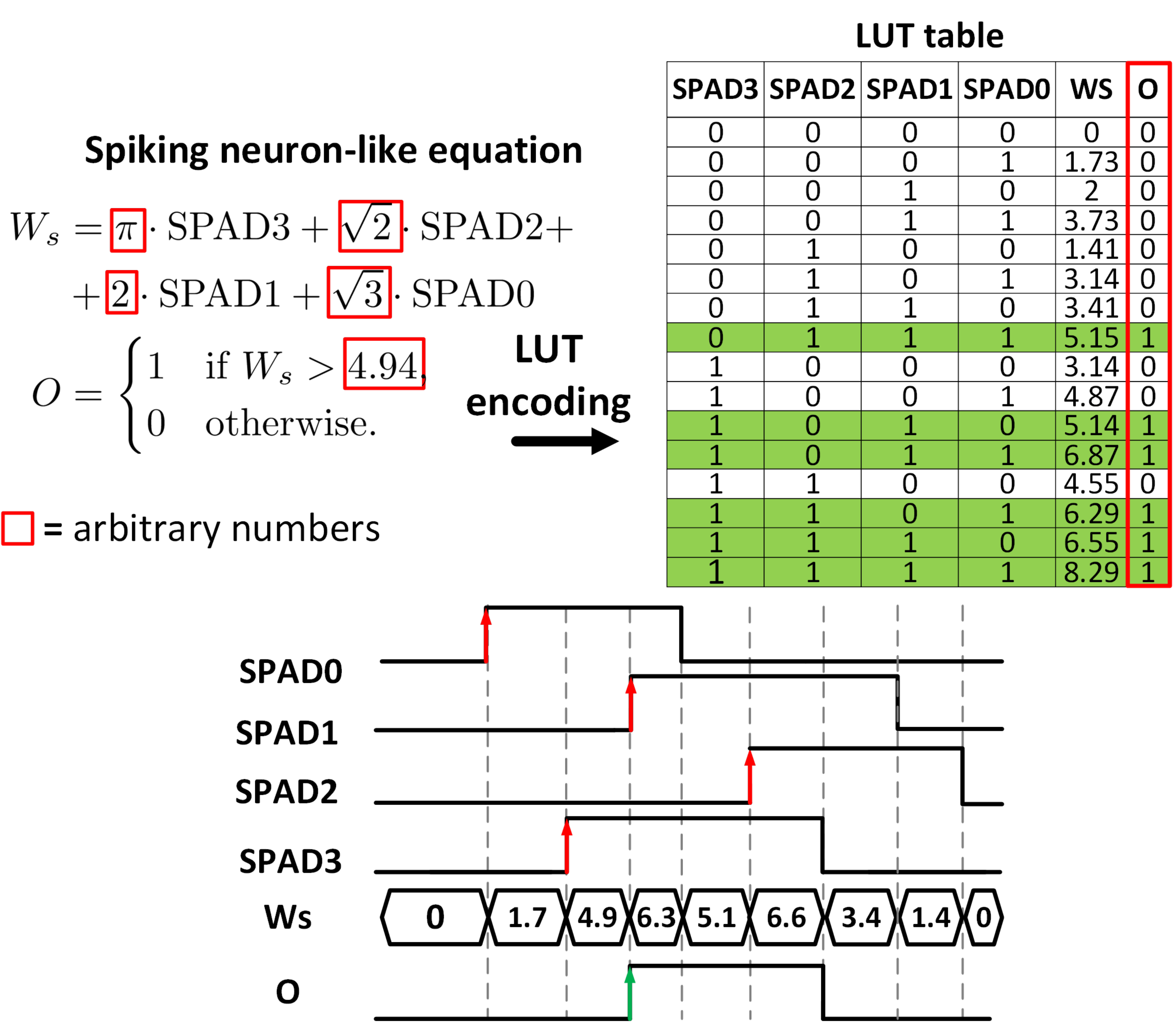}
    \caption{The weighted sum (WS) with the non-linear thresholding can be encoded in the LUT, which performs the operation in continuous time.}
    \label{fig_neuro_concept}
\end{figure}

By reading the SPAD through an LUT, it is possible to perform mathematical operations based on the digital pulse width and arrival time. A practical example is shown in Fig.~\ref{fig_neuro_concept}.
\vspace{-0.1in}
\subsection{Results}
\begin{figure}[h!]
    \centering
    \includegraphics[trim= 1 30 1 1,clip,width=0.8\linewidth]{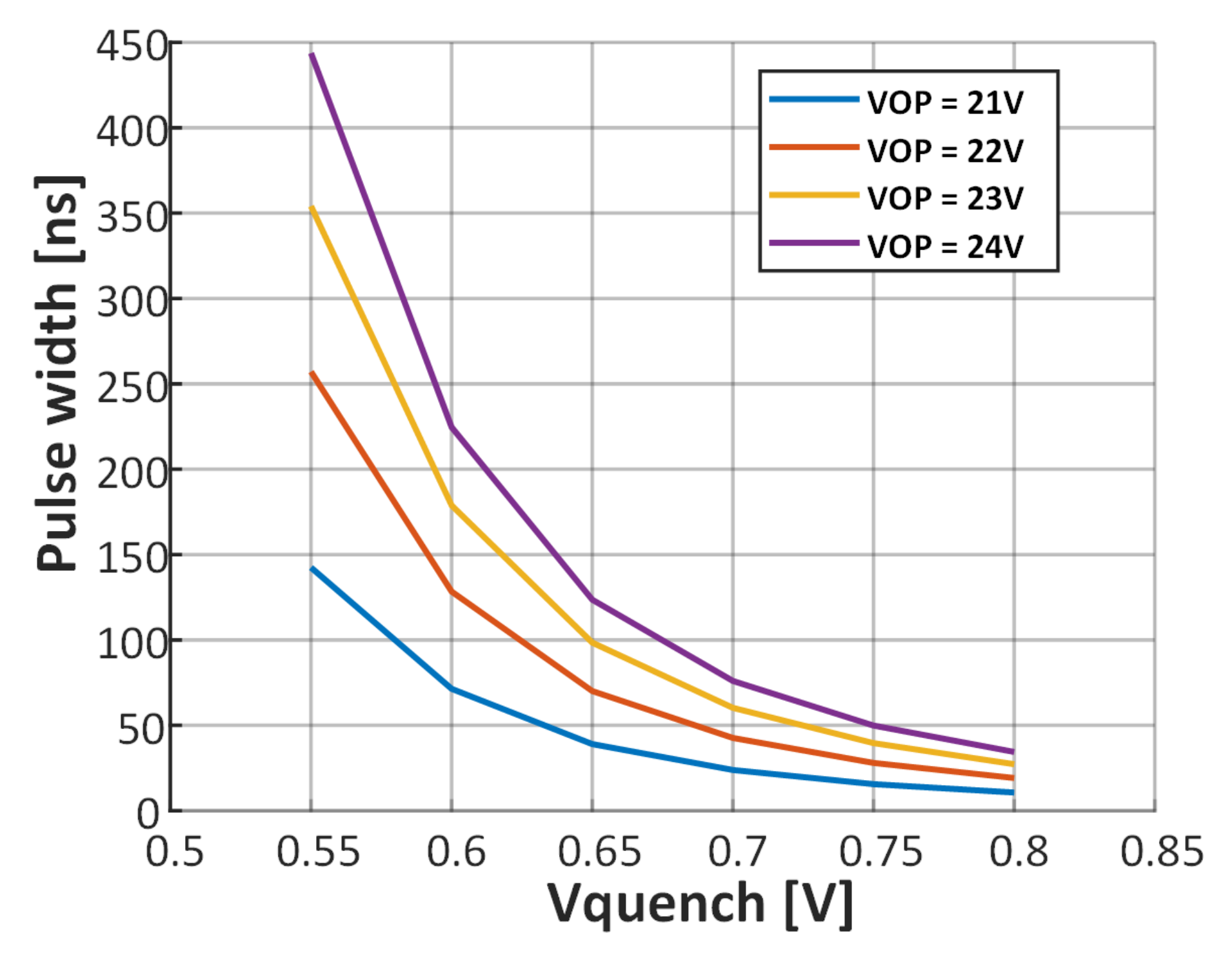}
    \caption{Pulse width of a pixel for different quenching voltages and SPAD bias. The breakdown voltage has been observed at $\approx 20\,V$. The configuration used to obtain this result is shown in Fig.~\ref{fig_SPADcombexample} (a) or (b).}
    \label{fig_DCR_mpix}
\end{figure}

\begin{figure}[h!]
    \centering
    \includegraphics[trim= 10 30 1 1,clip,width=0.8\linewidth]{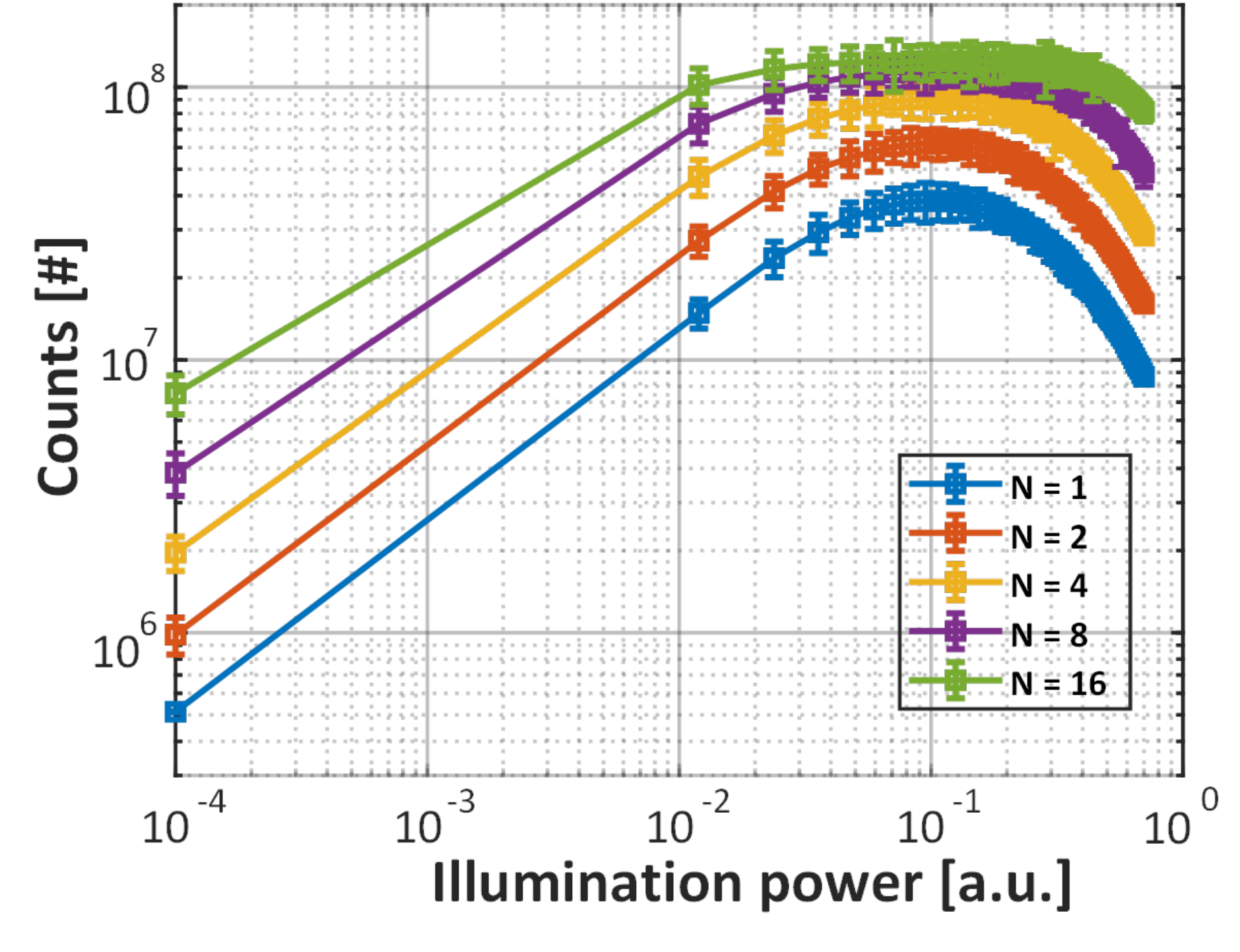}
    \caption{Linearity measurement with the XOR combinations by varying the number of SPADs connected to the XOR tree made with LUTs. The rising edges of the LUT output are counted over an integration time of 1 s and the mean over $100$ measurements is plotted. Right: configuration used to achieve this result. The configuration used to obtain this result is shown in Fig.\ref{fig_SPADcombexample} (c), adapted to 5 LUTs.}
    \vspace{-0.1in}
    \label{fig_linearity_XORcomb}
\end{figure}

\begin{figure}[h!]
    \centering
    \includegraphics[trim= 10 30 1 1,clip,width=0.8\linewidth]{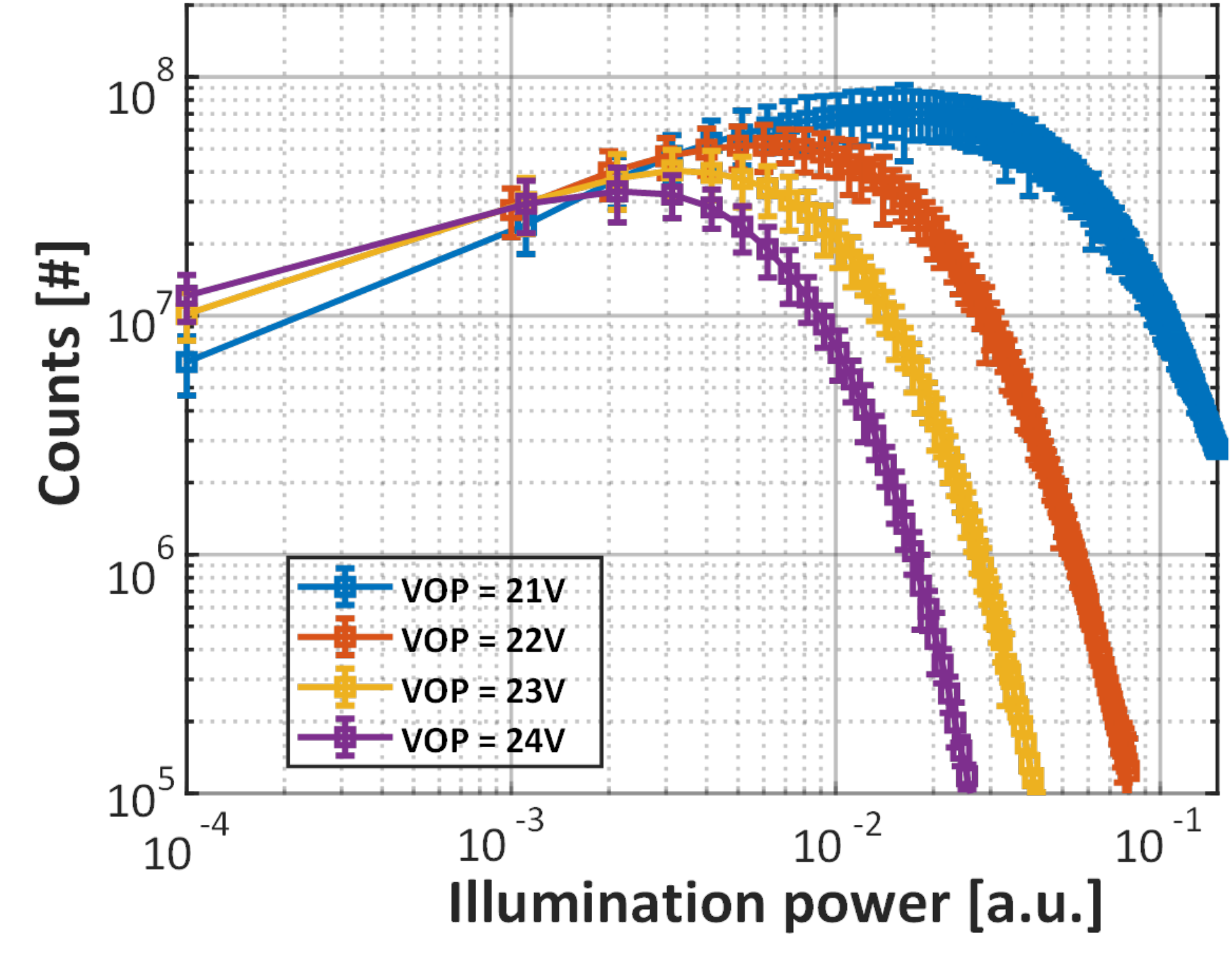}
    \caption{Linearity measurement with all the 16 SPADs connected to an OR tree implemented with LUTs. Considering the passive quenching and recharge circuit of Fig.~\ref{fig_pix_circuit}, at low illumination a higher SPAD bias leads to better detection efficiency, while at higher flux it entails a longer dead time and thus a reduced maximum count rate. The configuration used to obtain this result is shown in Fig.\ref{fig_SPADcombexample} (d), adapted to 5 LUTs.}
    \vspace{-0.2in}
    \label{fig_linearity_ORcomb}
\end{figure}

The pulse width plot is shown in Fig.~\ref{fig_DCR_mpix}.
Further characterizations of linearity over illumination power are shown in Fig.~\ref{fig_linearity_XORcomb} and Fig.~\ref{fig_linearity_ORcomb}, where for each graph a reference to the chosen configuration of Fig.~\ref{fig_SPADcombexample} is specified.

\section{Readout architecture}
\begin{figure*}[h!]
    \centering
    \includegraphics[trim= 5 20 1 10,clip,width=0.65\linewidth]{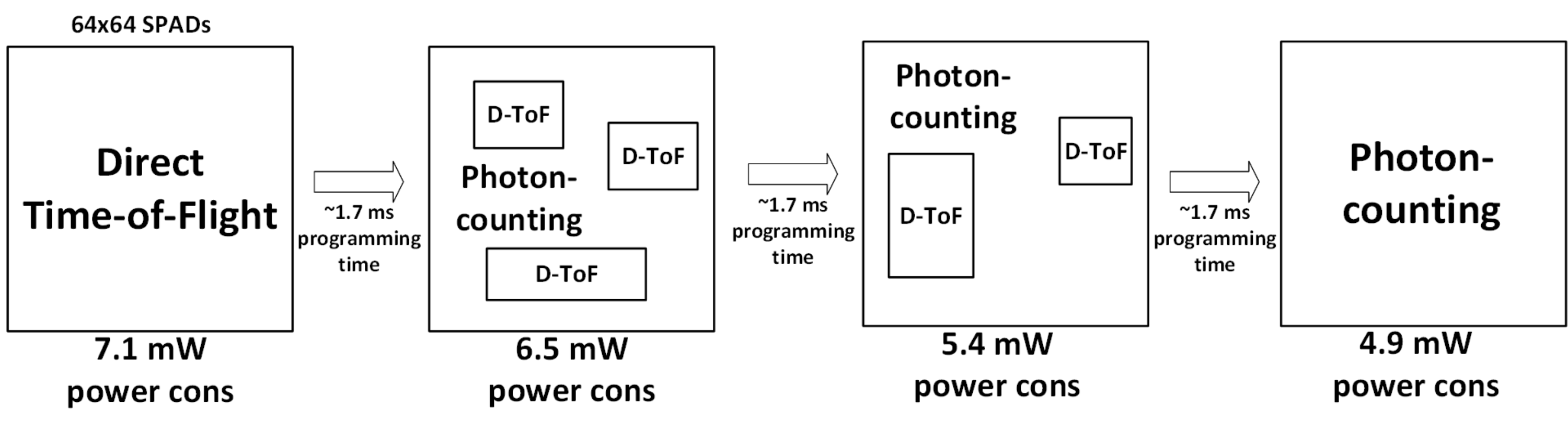}
    \caption{Power reduction by selectively configuring the array to be either operating in Time-of-Flight and/or in intensity mode. The power consumptions figures are based on post-layout simulation of a single macropixel at the typical corner and scaled to the target array size, i.e., multiplied by 1024.}
  
    \label{fig_power_reduction array}
\end{figure*}

\begin{figure}[h!]
    \centering
    \includegraphics[width=0.9\linewidth]{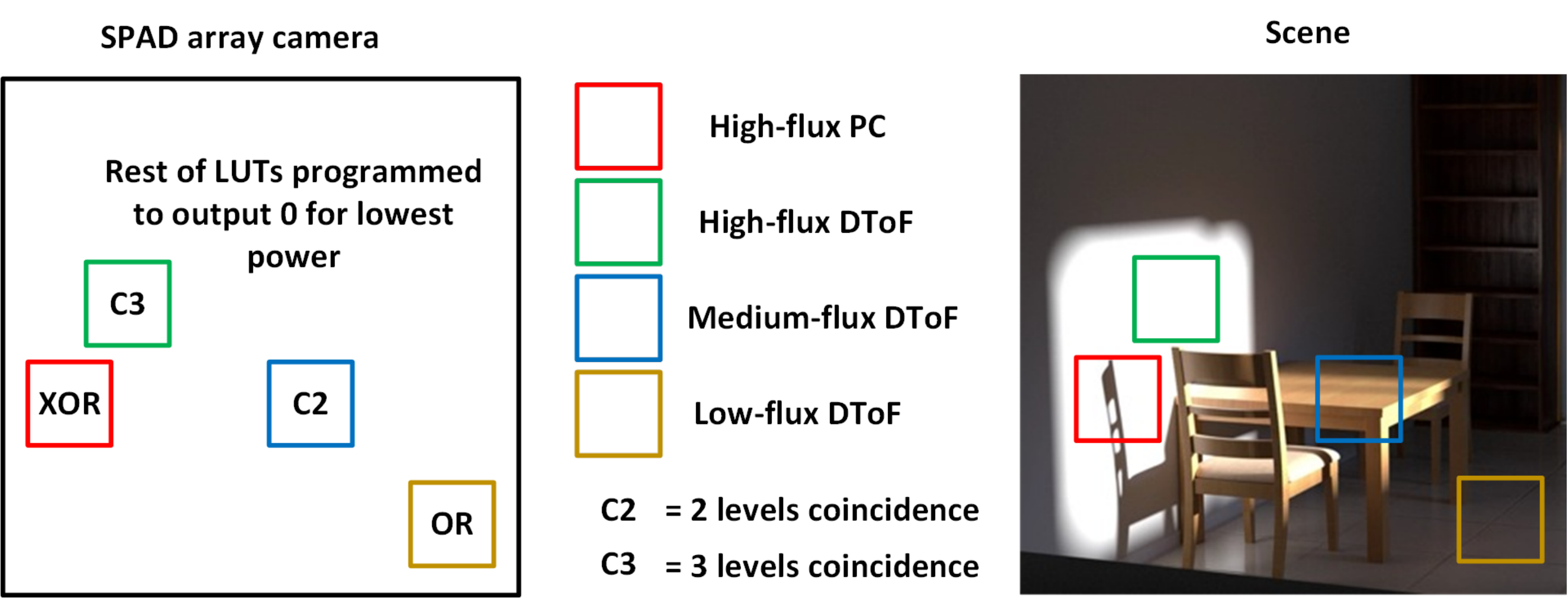}
    \caption{SPAD readout for scene adaptability. The PC part is useful for live background estimation. Three-level coincidence would be useful to filter background light which in a first-photon TDC architecture would lead to histogram distortion and low SBR.}
    \vspace{-0.2in}
    \label{fig_reconfigurable_fov_coincidence}
\end{figure}

\begin{figure}[h!]
    \centering
    \includegraphics[trim= 1 1 1 1,clip,width=0.75\linewidth]{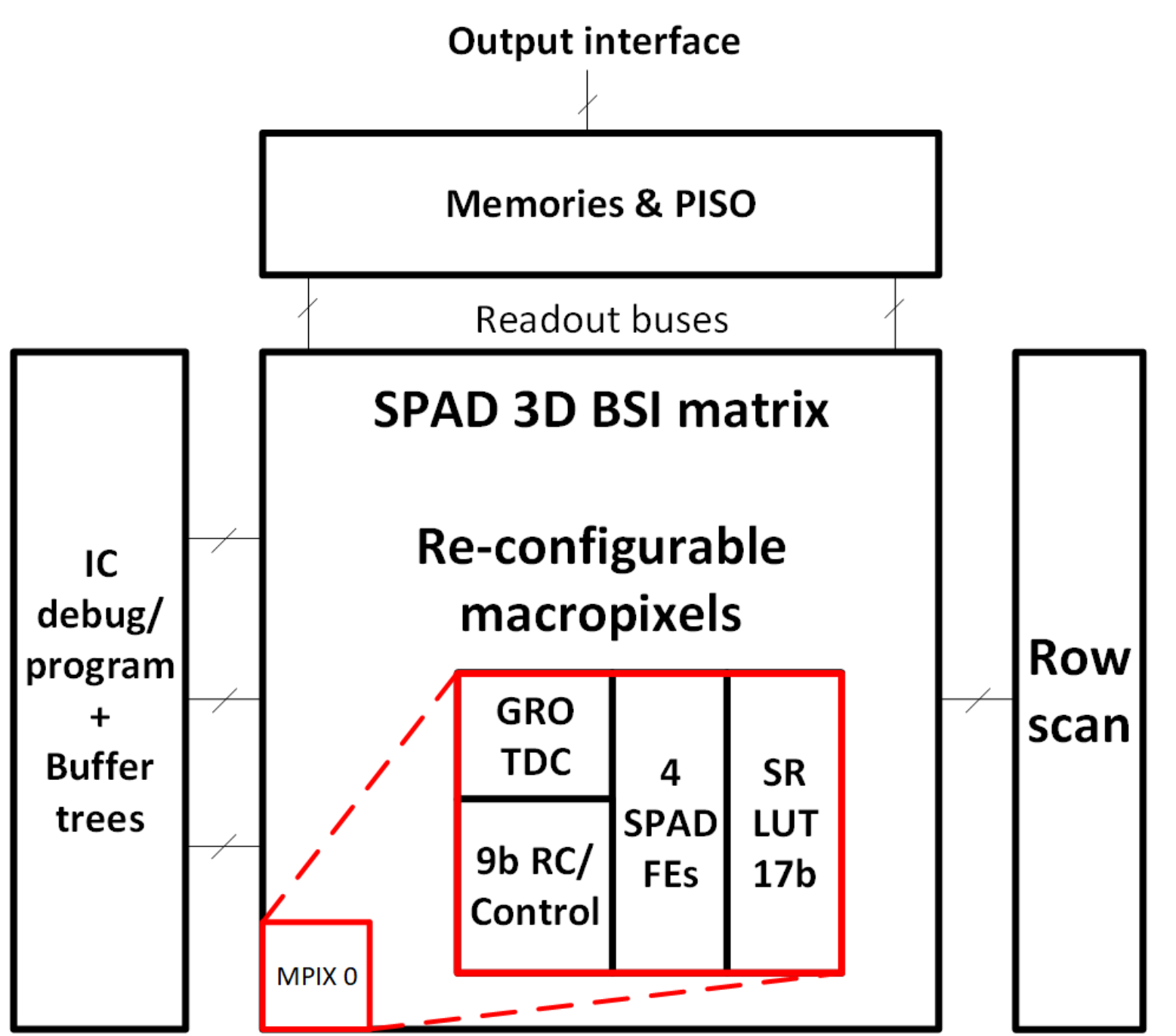}
    \caption{IC block diagram with inset on a macropixel floorplan.}
    \vspace{-0.2in}
    \label{fig_kelpie_block_diagram}
\end{figure}

The output of the macropixel $\mathrm{MPIX\_OUT}(i)$ of Fig.~\ref{fig_system_FPGA_archi} can be reorganized to enable the detection of ToF and photon counting intensity. Cluster outputs are reconfigured and mixed in an arbitrary way, as shown in Fig.~\ref{fig_power_reduction array}. Since photon-counting intensity requires less power than ToF, it becomes possible to trade functionality with power dissipation. In addition, this technique can reduce data rate quite significantly, while introducing a photon-driven approach to the readout. 
Each macropixel can also be reconfigured in terms of SPAD readout, adapting the SPAD combinations according to the scene FoV images by the macropixel. The corresponding approach is shown in Fig.~\ref{fig_reconfigurable_fov_coincidence}. A block diagram of the full IC is shown in Fig.~\ref{fig_kelpie_block_diagram}; it comprises a macropixel reconfiguration section and a global sensor reconfiguration section, both controlled independently.

\subsection{Programmable macropixel}
The macropixel comprises 2x2 SPADs ideally implemented in 3D-stacked technology. The pixel quenches the SPAD on the top tier and sends charge pulses through a floating capacitance to the bottom-tier circuit. The SPAD is gated with a NAND and the digital pulse is forwarded to the LUT.
The physical dimensions are fixed by the SPAD pitch of 10.17 $\mu\mathrm{m}$, leading to 20.34 $\mu\mathrm{m}$ pitch. Upon photon detection, SPAD-generated digital pulses are sent to the LUT, which performs a spatio-temporal compression and sends the pulse to the control logic.
A 17 bits shift register is used to program the macropixel, 16 bits for the LUT and 1 bit to choose between photon-counting or ToF operating modes. The output of the LUT signal is sent to the control logic, responsible for the generation of a timing window if the macropixel is programmed as dToF. The TDC architecture enables reconfiguring the ripple counter to count the LUT pulses, thus enabling the photon-counting operation.
The macropixels are abutted along with their shift registers (SRs), leading to a 17408 SR to program all the 1024 macropixels with a 10 MHz clock, resulting in $\approx\,1.7 \mathrm{ms}$ of programming time,enough for standard physical scenes. By choosing the SR approach, the macropixel layout area and the programming time are increased with respect to other memory architectures like SRAM cells or latches. However, easy testing is achieved.

\subsection{Time-to-digital converter}
\begin{figure}[h!]
    \centering
    \includegraphics[trim= 1 1 1 10,clip,width=1\linewidth]{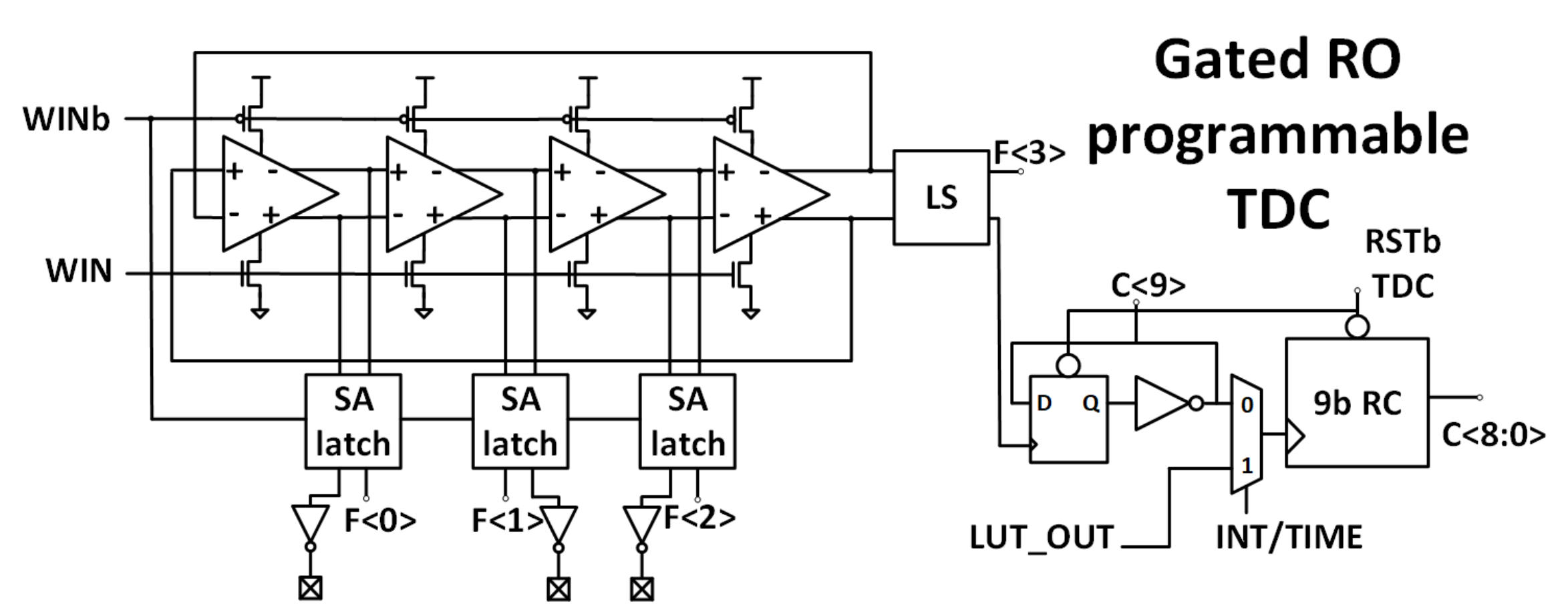}
    \caption{TDC schematics. The GRO has also a reset signal to be able to make it start in a known state. The WIN signal is generated only if the macropixel is programmed for ToF operation.}
    \vspace{-0.1in}
    \label{fig_TDC_archi}
\end{figure}

The TDC is based on a differential gated ring oscillator (GRO) similar to \cite{veerappan2011160}. The 4 GRO phases are sensed by strong-arm latches implemented with low-threshold transistors and sent to tri-state buffers over a column line. One GRO phase is sent to a level shifter (LS) and subsequently to a first toggle flip-flop and then to a ripple counter (RC), employing an architecture similar to \cite{henderson2019192}. The asynchronous RC counts the GRO oscillations during its active time. The final timing data comprises 14b, 4b fine from the GRO and 10b coarse from the toggle FF and the RC. A schematic of the TDC is shown in Fig.~\ref{fig_TDC_archi}.

\section{Conclusions \& Future work}
In this paper, we have introduced an FPGA underneath SPAD pixels and we have proposed a reconfigurable architecture to selectively read SPADs in two aquisition modes that can be mixed in the array. This approach enables arbitrary configurations thanks to the digital nature of the SPADs, which are organized as macropixels, thus potentially reducing data rate and power, while
improving background light estimation. 

\section*{Acknowledgments}
\noindent
Tommaso Milanese’s PhD is funded by STMicroelectronics.

\bibliographystyle{IEEEtran}
\bibliography{references}

\end{document}